\documentclass{article}

\usepackage{arxiv}
\usepackage[utf8]{inputenc} 
\usepackage[T1]{fontenc}    
\usepackage{hyperref}       
\usepackage{url}            
\usepackage{booktabs}       
\usepackage{amsfonts}       
\usepackage{nicefrac}       
\usepackage{microtype}      
\usepackage{lipsum}
\usepackage{graphicx}
\graphicspath{ {./images/} }
\usepackage{graphicx}
\usepackage{lineno,hyperref}
\usepackage{booktabs} 
\usepackage{amsmath}
\usepackage{bbm}
\usepackage{amssymb,amsfonts}
\usepackage{algorithm}
\usepackage[noend]{algpseudocode}
\usepackage{tabularx}
\usepackage{multirow}
\usepackage{siunitx}
\usepackage{amsmath}

\makeatletter
\def\algbackskip{\hskip-\ALG@thistlm}
\makeatother
\modulolinenumbers[5]

\title{SumRecom: A Personalized Summarization Approach by Learning from Users' Feedback}

\author{
  Samira Ghodratnama \\
  Macquarie University\\
  Sydney, Australia\\
  \texttt{samira.ghodratnama@mq.edu.au} \\
   \And
  Mehrdad Zakershahrak \\
  Macquarie University\\
  Sydney, Australia\\
  \texttt{mehrdad.zakershahrak@mq.edu.au} \\
}

\begin{document}
\maketitle
\begin{abstract}
Existing multi-document summarization approaches produce a uniform summary for all users without considering individuals' interests, which is highly impractical. 
Making a user-specific summary is a challenging task as it requires: i) acquiring relevant information about a user; ii) aggregating and integrating the information into a user-model; and iii) utilizing the provided information in making the personalized summary.
Therefore, in this paper, we propose a solution to a substantial and challenging problem in summarization, i.e., recommending a summary for a specific user.
The proposed approach, called \textit{SumRecom}, brings the human into the loop and focuses on three aspects: personalization, interaction, and learning user's interest without the need for reference summaries.
\textit{SumRecom} has two steps: i) The user preference extractor to capture users' inclination in choosing essential concepts, and ii) The summarizer to discover the user's best-fitted summary based on the given feedback.
Various automatic and human evaluations on the benchmark dataset demonstrate the supremacy \textit{SumRecom} in generating user-specific summaries. 
\end{abstract}

\section{Introduction}
\label{intro}
Document summarization is used to extract the most informative parts of documents as a compressed version for a particular user or task~\cite{ghodratnama2023personalized,beheshti2021bp}.
A good summary should keep the fundamental concepts while helping users to understand large volumes of information quickly.
However, it is still hard to produce summaries that are comparable with human-written ones~\cite{ghodratnama2021intelligent,khanna2022transformer}.
A significant challenge is a high degree of subjectivity in content selection, i.e., what is considered essential for different users.
Optimizing a system towards one best summary that fits all users is highly impractical as it is the current state-of-the-art.
Making a user-specific summary for an input document cluster is a challenging task. 
It requires: i) acquiring relevant information about a user, ii) aggregating and integrating the information into a user-model, and iii) using the provided information in making the personalized summary.
We bring the human in the loop and create a personalized summary that better captures the users' needs and their different notions of importance. 
Our rationale behind this is:
\begin{itemize}
    \item Keeping humans in the loop by giving feedback through interaction.
    Besides, to reduce users' cognitive burden in giving feedback, we consider two aspects.
    First, feedback is given in the form of preference.
    Second, the preference is in the form of concepts, not a complete summary.
    Moreover, users are allowed to define the detailed properties of produced summaries.
    It also helps in reducing the search space by leveraging given feedback in making summary space.
    \item Evaluating the quality of a summary based on a domain expert according to the user's feedback.
    A learner must understand how to generate an optimal summary for a user based on the evaluation metric.
\end{itemize}
We summarize our contribution in three aspects: personalization, interaction, and learning user's interest without the need for reference summaries, as discussed below.

\subsection{Personalization}
Existing multi-document summarization approaches produce a uniform summary for all users without considering individuals' interests.
Therefore, summaries are not interpretable and personalized.
They are also designed to create short summaries and incapable of producing more extended summaries.
Therefore, all details are omitted even if the user is interested in more information.
Unfortunately, a single summary is unlikely to serve all users in a large population.
Therefore, a good summary should reflect users' preferences and interests.
As a result, a good summary should change per the preferences of its reader.
Therefore, summaries need to reflect users' interests and background in making summaries.
For instance, there is various information available on the internet about COVID-19.
While one might be interested in symptoms, the other could be looking for the outbreak locations, while others are searching about the death toll.

\subsection{Interaction}
In a personalized approach, the system needs to know about the user's background knowledge or interests.
When we do not have access to the user's preferences and interests, including profile or background knowledge, the system requires interaction with users to acquire feedback for modeling users' interests.
Multiple forms of feedback have been studied for different applications, such as clicks, post-edits, annotations over spans of text, and preferences over pairs of outputs~\cite{ghodratnama2020adaptive,ghodratnama2020rare,ghodratnama2021summary2vec}. 

\subsection{Reference Summaries}
Most existing document summarization techniques require access to reference summaries made by humans to train their systems.
Therefore, it is costly and time-consuming.
A report shows that 3,000 hours of human efforts were required for a simple evaluation of the summaries for the Document Understanding Conferences (DUC)~\cite{lin2004rouge}.
Personalized summaries eliminate the need for reference summaries as they make a specific summary for a user instead of optimizing a summary for all users.
\begin{figure*}
    \centerline{\includegraphics[width=\textwidth]{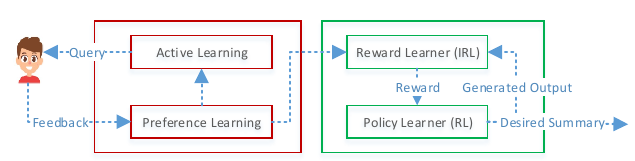}}
    \caption{An overview of the \textit{SumRecom} Approach: Active and preference learning are used to extract user's preferences.
    The learned preference ranking function is used to produce the desired summary using inverse reinforcement learning (IRL) for learning the reward and reinforcement learning for learning the optimal policy (RL).}
    \label{overview}
\end{figure*}
\subsection{Our contribution}
The proposed method, \textit{SumRecom} is a preference-based interactive summarization approach that extracts users' interests to generate user-adapted results.
\textit{SumRecom} predicts users' desired summaries by incrementally adapting the underlying model through interacting with users.
The proposed approach has two steps: i) The user preference extractor and ii) The summarizer.
Our model employs \textit{active learning} and \textit{preference learning} to extract users' preference in selecting contents. 
\textit{SumRecom} also utilizes \textit{integer linear programming (ILP)} to maximize user-desired content selection based on the given feedback.
It then proposes an \textit{inverse reinforcement learning (IRL)} algorithm and using the domain expert's knowledge for evaluating the quality of summaries based on the given feedback.
The learned reward function is used to learn the optimal policy to produce the user's desired summary using \textit{reinforcement learning (RL)}.
A general overview of the algorithm is depicted in Figure~\ref{overview} for more clarification.

The rest of the paper is organized as follows: in Section~\ref{Related}, we present the related work.
Section~\ref{proposed} discusses the methodology behind the personalized summarization approach.
In Section~\ref{Evaluation}, we present evaluation results before concluding the paper with remarks for future directions in Section~\ref{conclusion}.

\section{Related Work}
\label{Related}
To generate a summary, an agent requires the ability of natural language processing and background knowledge.
Thus, this process is complicated even for a domain-expert, yet it can be even more difficult for machines.
Early work on document summarization focused on a single document.
However, multi-document summarization gains more attention recently due to the massive development of documents~\cite{gupta2010survey}.
In the following, we discuss multi-document summarization approaches as it is the focus of this paper.
We categorize multi-document summarization approaches as traditional approaches, personalized and interactive approaches, and preference-based and reinforcement-learning-based approaches.

\subsection{Traditional Approaches}
The main category considers the process and the output type of the summarization algorithm: extractive and abstractive approaches. 
Abstractive summaries are generated by interpreting the main concepts of documents and then stating those contents in another format~\cite{li2020text}.
Abstraction techniques are a substitute for the original documents rather than a part of them.
Therefore, abstractive approaches require deep natural language processing, such as semantic representation and inference.
However, abstractive summaries are challenging to produce~\cite{mehta2018effective}.

Extractive multi-document summarization (EMDS) has been widely studied in the past.
Since the proposed approach in this paper is extractive, we analyze the extractive methods in more detail.
Given a cluster of documents on the same topic as input, an EMDS system needs to extract basic sentences from the input documents to generate a summary complying with a given length requirement that fits the user's needs~\cite{mehta2018effective}.
Early extractive approaches focused on shallow features, employing graph structure, or extracting the semantically related words~\cite{edmundson1969new}. 
Different machine learning approaches, such as naive-Bayes, decision trees, log-linear, and hidden Markov models are also used for this purpose~\cite{ghodratnama2015efficient,ghodratnama2020extractive}.

Recently, the focus for both extractive and abstractive approaches is mainly on neural network-based and deep-reinforcement learning methods, which could demonstrate promising results.
They employ word embedding~\cite{pennington2014glove} to represent words at the input level.
Then, feed this information to the network to gain the output summary. 
These models mainly use a convolutional neural network~\cite{cao2015learning}, a recurrent neural network~\cite{cheng2016neural,nallapati2017summarunner} or the combination of these two~\cite{wu2018learning,narayan2018ranking}.
JECS~\cite{xu2019neural}, PGN~\cite{seeget}, and DCA~\cite{celikyilmaz2018deep} are recent abstractive state-of-the-art approaches.
JECS is a neural text-compression-based summarization approach that uses BLSTM as the encoder.
It first selects sentences and then prunes the parsing tree to compress chosen sentences.
PGN is a pointer generator network that works based on encoder-decoder architecture.
DCA (Deep Communicating Agents) works based on the hierarchical attention mechanism.

In a recent attempt, they used BERT, a pre-trained transformer to produce extractive summaries, called \textit{BERTSUM}~\cite{liu2019fine}.
\textit{BERTSUM} has achieved ground-breaking performance on multiple NLP tasks.
HIBERT~\cite{zhang2019hibert}, PNBERT\cite{zhong2019searching}, BertSumExt, and BertSumAbs~\cite{liu2019text} are also recent state-of-the-art BERT-based approaches.

\subsection{Personalized and Interactive Approaches}
There exist few recent attempts on personalized and interactive approaches in various NLP tasks.
Unlike non-interactive systems that only present the system output to the end-user, interactive NLP algorithms ask the user to provide certain feedback forms to refine the model and generate higher-quality outputs tailored to the user.
Multiple forms of feedback also have been studied including mouse-clicks for information retrieval~\cite{borisov2018click}, post-edits and ratings for machine translation~\cite{denkowski2014learning,kreutzer2018can}, error markings for semantic parsing~\cite{lawrence2018counterfactual}, and preferences for translation~\cite{kingma2014adam}. 

In the summarization task, most existing computer-assisted summarization tools present important elements of a document or the output of a given automatic summarization system to the user.
The output is a summary draft where they ask users to refine the results without further interaction.
The refined process include to cut, paste, and reorganize the important elements to formulate a final text~\cite{orasan2006computer,craven2000abstracts,narita2002web}.

Other works present automatically derived hierarchically ordered summaries allowing users to drill down from a general overview to detailed information~\cite{christensen2014hierarchical,shapira2017interactive}.
Therefore, these systems are neither interactive nor consider the user's feedback to update their internal summarization models.
Other interactive summarization systems include the iNeATS~\cite{leuski2003ineats} and IDS~\cite{jones2002interactive} systems that allow users to tune several parameters (e.g., size, redundancy, focus) for customizing the produced summaries.
Avinesh and Meyer~\cite{avinesh2017joint} proposed a most recent interactive summarization approach that asks users to label important bigrams within candidate summaries.
Their system can achieve near-optimal performance in ten rounds of interaction in simulation experiments, collecting up to 350 critical bigrams.
However, labeling important bigrams is an enormous burden on the users, as users have to read through many potentially unimportant bigrams.

Considering the interactivity structure of the proposed approach, it can also be comparable with query-focused summarization approaches.
In query-based summarization, a good summary is relevant to a specific query.
One common technique in this category is to adapt existing summarization approaches for answering a query.
In this way, the importance of a sentence should be judged by its significance to the user’s query.
Using topic signature words or graph-based approaches~\cite{qazvinian2010citation}, and sub-modular approaches~\cite{lin2011class,xu2020coarse} are examples of this category.
Other approaches also exist which are explicitly designed for answering the query~\cite{abdi2018qmos}.
Recent studies include Dual-CES~\cite{roitman2020unsupervised}, a novel unsupervised approach that uses the Cross-Entropy Summarizer (CES).
The focus is to handle the tradeoff between saliency and focus in summarization.
QMDS~\cite{pasunuru2021data} is another study in the abstractive query-focused summarization approach where they tackle two problems. 
First, generating large-scale, high-quality training datasets using data augmentation techniques.
Second, they build abstractive end-to-end neural network
models on the combined datasets that yield new state-of-the-
art transfer results on DUC datasets. 

Evaluating interactive summarization approaches is another challenge.
In a study~\cite{shapira2021extending}, authors developed  an end-to-end evaluation framework, focusing on expansion-based interaction.
It considers the accumulating information along a user session. 

\subsection{Preference-based and Reinforcement-based Approaches}
There is an increasing research interest in using preference-based feedback and reinforcement learning algorithms in summarization.
As an example, one approach is to learn a sentence ranker from human preferences on sentence pairs in ~\cite{zopf2018estimating}.
The ranker then is used to evaluate the quality of summaries by counting the number of high-ranked sentences included in a summary. 
Reinforcement-learning-based (RL) approaches are another popular category for both extractive and abstractive summarization in recent years~\cite{ryang2012framework,pasunuru2018multi,paulus2018deep}.
Most existing RL-based document summarization systems use heuristic functions as the reward function and, therefore, do not rely on reference summaries~\cite{ryang2012framework,rioux2014fear}.
Some other approaches use different ROUGE measure variants as the reward function and therefore require reference summaries as the rewards for RL~\cite{pasunuru2018multi,paulus2018deep,kryscinski2018improving}. 
However, neither ROUGE nor the heuristics-based rewards can precisely reflect real user's preferences on summaries~\cite{chaganty2018price}.
Therefore, using these imprecise reward models can severely mislead the RL-based summarizer. 
One challenge in RL-based summarization approaches is defining the reward function~\cite{gao2019reward}.

\begin{figure*}[t]
    \centerline{\includegraphics[width=\textwidth]{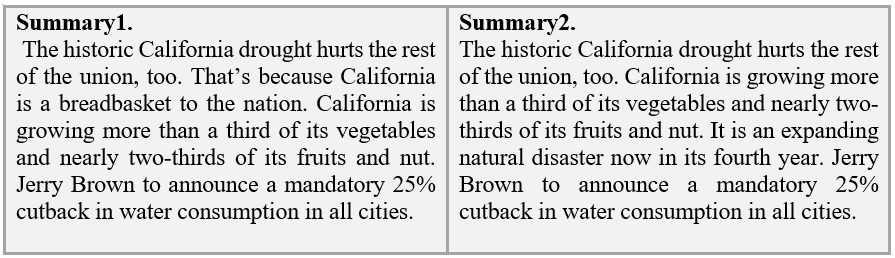}}
    \caption{Example of comparing two summaries that put a substantial cognitive burden on users.}
    \label{comparing}
\end{figure*}

The preference-based and reinforcement learning algorithms also have been used in summarization simultaneously. 
The first approach is SPPI~\cite{sokolov2016stochastic,kreutzer2017bandit}, a policy-gradient RL algorithm that receives rewards from the preference-based feedback.
The problem is that SPPI suffers heavily from the high sample complexity problem.
Another recent preference reinforcement learning approach is APRIL~\cite{gao2018april}, which has two stages.
First, the user’s ranking over candidate summaries is retrieved, and then a neural reinforcement learning (RL) agent is used to search for the optimal summary.
However, preferring one summary over the other one in both approaches puts a considerable burden on users.
It is worth mentioning that summarization aims to provide users with a summary, which helps them not read numbers of documents.
While asking users to prefer a summary over another in multiple rounds among a summary space that includes all randomly possible combinations of sentences puts an additional cognitive load on them, that is even more than reading the documents.
Figure~\ref{comparing} represents an example of this comparison, where it can be even more challenging when the length of the summary increases.

\section{Proposed Method (SumRecom)}
\label{proposed}
One of the ultimate goals of machine learning is to provide predictability for unseen situations. 
Therefore, personalizing the summaries is one of the fundamental approaches to construct summaries tailored to the user's demands.
This paper proposes a human-in-the-loop approach to create a personalized summary that better captures the users' needs.
Making a user-specific summary for an input document cluster is a challenging task.
The system should have background knowledge about the user to be able to produce personalized summaries.
Since we do not consider any prior knowledge, including user profile or background knowledge, the system requires user interaction to model user interest.

\begin{figure*}[t]
    \centerline{\includegraphics[width=\textwidth]{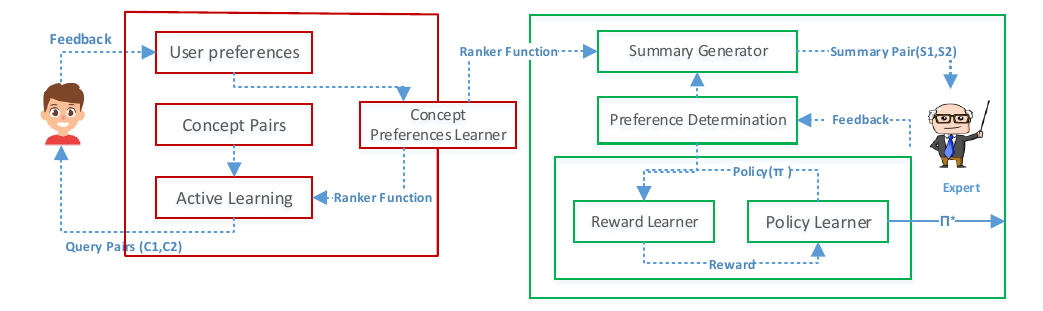}}
    \caption{\textit{SumRecom} approach in more detail: 1) The left side is the user preference extractor using active preference learning over concepts, and 2) the right side is the summarizer including reward learning (IRL) and policy learning (RL).}
    \label{DetailedAlg}
\end{figure*}

\textit{SumRecom} considers the summarization problem as a recommender system where the goal is to suggest a personalized summary to a user based on the given preferences.
This novel framework has two components: i) The user preference extractor and ii) The summarizer.
The \textit{user preference extractor} is responsible for querying the user and potentially receiving the feedback using \textit{active preference learning}.
The summarizer aims to generate summaries based on users' feedback and learns how to make a user-desired summary by developing a reinforcement learning algorithm (IRL)~\cite{abbeel2004apprenticeship}.
The process is depicted in Figure~\ref{DetailedAlg}, and the overall algorithm is reported in Algorithm~\ref{alg1}.
\begin{algorithm}[t]
    \caption{SumRecom algorithms.}\label{alg1}
    \begin{flushleft}
    \hspace*{\algorithmicindent} \textbf{Input}: Document Cluster x. \\
    \hspace*{\algorithmicindent} \textbf{Output}: Optimal Summary for a user. 
    \end{flushleft}
    \begin{algorithmic}[1]
    \Procedure{SumRecom.}{}
    \State $\textit{Concepts} \gets \textit{Concept Extraction (x)}$
    \BState \emph{Modeling User Preference}:
    \State $\textit{Query pairs} \gets \textit{Active Learner(Concepts)}$
    \State $\textit{User Preferences} \gets \textit{Query pairs (user)}$
    \State $\textit{Ranker Function} \gets \textit{Preference Learner (User Preferences)}$
    \BState \emph{The summarizer}:
    \State $\textit{Summaries} \gets \textit{Summary Generator(Ranking  Function)}$
    \State $\textit{Summary Ranker} \gets \textit{Reward Learner(Summaries)}$
    \State $\textit{Optimal Policy} \gets \textit{Policy Learner(Summary Ranker)}$
    \EndProcedure
    \end{algorithmic}
\end{algorithm}

\subsection{The user preference extractor}
Understanding users' interests is the first step towards making personalized summaries.
Users' interest can be extracted implicitly based on users' profiles, browsing history, likes or dislikes, or retweeting in social media~\cite{alhindi2015profile}.
Consequently, interaction is an approach to predict user's perspectives in the new circumstances based on the feedback user provided in the past ~\cite{zakershahrak2018interactive,zakershahrak2020order}.
The user feedback can be in any form, such as mouse-click or post-edits.
Further, experiments suggest that preference-based interactive approaches put a lower cognitive burden on human subjects than asking for absolute ratings or categorized labels as it is a binary decision~\cite{zopf2018estimating,kingsley2010preference}.
Besides, preferring one summary over another puts a significant burden on the user, as discussed in section~\ref{intro}.
For instance, when collecting feedback about a user's interest, asking the user to compare the concepts "cancer treatment" and "cancer symptoms" involves a smaller cognitive workload than asking the user to assign a score to each of the concepts.
On the other hand, it is challenging for users to decide the usefulness of a summary throughout a scoring scheme.
Therefore, in this paper, to reduce users' cognitive load, queries are in the form of concept selection, and the feedback is in the form of preferences. 

Concept selection aims to find the critical information within a given set of source documents as humans can quickly assess the importance of concepts given a topic.
Since the notion of importance is specific to a particular topic or user, we query users to ask preference over concepts.
Users can better prefer one concept to the other instead of selecting the important concept.
However, to collect enough data to make a meaningful conclusion, it is required for users to interact with the system in many rounds to simulate the ideal user feedback.
Therefore active learning is also used to reduce the number of interaction rounds.
To recap, we use active preference learning (APL) in an interaction loop to maximize the information gained from a small number of preferences, reducing the sample complexity.
In the following, the active and preference learning implemented in the proposed method is discussed.

\subsubsection{Preference Learning}
Preference learning is a classification method that learns how to rank instances based on observed preference information.
In other words, it trains based on a set of pairwise preferred items and obtaining the total ranking of objects ~\cite{furnkranz2010preference}.

To formally define the preference learning in our proposed algorithm, let $X$ be the input space and $x$ a cluster of documents. 
Let's define $C(x)$ all extracted concepts from document cluster $x$.
Therefore, we have a concept space $C(x)=\{c_1,c_2,..,c_N\}$ with $N$ concepts.
The goal is to query users a set of pairwise preference of concepts $\{p(c_{11},c_{21}),p(c_{12},c_{22}),...,p(c_{1n},c_{2n})\}$ where $p(c_{1i},c_{2i})$ is a preference instance showing to user in $i-th$ round where:
\begin{equation}
  p(c_{1i},c_{2i})=\begin{cases}
    1, & \text{if $c_{1i}>c_{2i}$}.\\
    0, & \text{otherwise}.
  \end{cases}
\end{equation}

where $>$ indicates the preference of $c_{1i}$ over $c_{2i}$.
Then, the goal of preference learning is to predict the overall ranking of concepts.
If we can find a mapping from data to real numbers, ranking the data can be solved by ranking the real numbers.
This mapping is called utility function $U$ such that
$ c_i > c_j \xrightarrow{} U(c_i) > U (c_j)$ where $U$ is a function $U: C\xrightarrow{}  \mathbbm{R}$.

In this problem, the ground-truth utility function ($U$) measures each concept's importance based on users' attitudes.
We assume that no two items in $C(x)$ have the same $U$ value which is a required condition for learning process.
Finding the utility function is a regression learning problem that is well-studied in machine learning. 
In this problem, the ranking function ($R$) measures the importance of each concept based on users' attitude toward other concepts defined as:

\begin{equation}
    R(c_i)=\sum \mathbbm{1} \{U(c_i)>U(c_j)\} , \forall c_j \in {C(x)}
\end{equation}
where $\mathbbm{1}$ is the indicator function. 
Therefore, $R$ gives the rank of $c_i$ among all extracted in $x$ ($C_x$).

The Bradley–Terry model~\cite{szummer2011semi} is a probability model widely used in preference learning.
Given a pair of individuals $c_i$ and $c_j$ drawn from some population, it estimates the probability that the pairwise comparison $c_i > c_j$ turns out true.
Having $n$ observed preference items, the model approximates the ranking $R$ by computing maximum likelihood estimate:

\begin{equation} \label{objectiveTR}
\begin{split}
J_x(w)= & \sum_{i \in n}[p(c_{1i},c_{2i})\log H(c_{1i},c_{2i};w)+\\
 &  [p(c_{2i},c_{1i})\log H(c_{2i},c_{1i};w)))]
\end{split}
\end{equation}

where  $H(c)$ is the logistic function defined as:

\begin{equation} \label{eq2}
 H(c_{i},c_{j};w)= \frac{1}{1+exp [U^{*}{(c_j;w)}-U^{*}{(c_i;w)}]}
\end{equation}

and $U^{*}$ is the approximation of $U$ parameterised by $w$ which can be learned by different function approximation techniques.
In our problem a liner regression model is designed for this purpose defined as:

\begin{equation} \label{eq3}
U(c;w)=w^{T}\phi(c)
\end{equation}

where $\phi(c)$ is the representation feature vector of concept $c$.
For any $c_i,c_j \in C$, the ranker prefers $c_i$ over $c_j$ if $w^{T}\phi(c_i)> w^{T}\phi(c_j)$.
By maximizing the $J_x(w)$ in Eq.~\ref{objectiveTR} we have:

\begin{equation} \label{eq31}
w^{*} = \operatorname*{arg\,max}_w J_x(w)
\end{equation}

The resulting $w^{*}$ using stochastic gradient ascent optimization will be used to obtain $U^{*}$, which in turn can be used to induce the approximated ranking function $R^{*}: C \xrightarrow[]{} \mathbbm{R}$.
Maximization of this objective function will assure those high probabilities are assigned to pairs with low loss.
\textit{SumRecom} learns a ranking over concepts and uses the ranking to guide the summarizer discussed in the following.

\subsubsection{Active Learning}
\label{sec:activelearn}
To emphasize the need for \textit{active learning}, let's consider we have $M$ sentences to summarize, and each sentence has $4$ unique concepts on average.
As a result, the number of unique concepts is  $4\times M$.
Therefore, the number of pairwise preferences to query the user to have a complete comparison in this setting is equal to $\binom {4M}2=\frac{4M!}{2!(4M-2)!}$.
As an example, if $M=100$, this number is equal to $79800$, which is impossible.
Therefore, active learning aims to find the minimum subset of best samples, in our problem, the best pairs, to query the user to gain the most information.
Therefore, the number of examples to query user is much lower than the number required in regular supervised learning. 
There exist different strategies to find the minimum subset of best samples.
Examples include~\cite{settles2009active}:
\label{strategies}
\begin{itemize}
    \item \textbf{Balance exploration and exploitation}: The exploration and the exploitation of the data space representation is the measure to choose samples.
    In this strategy, the active learning problem is modeled as a contextual bandit problem.
    \item \textbf{Expected model change}: The policy behind this model is to select the samples that would most change the current model.
    \item \textbf{Expected error or variance reduction}: This strategy selects samples that would most reduce the model's generalization error or variance.
    \item \textbf{Uncertainty sampling}: The idea is to select samples for which the current model is least certain to the correct output.
    \item \textbf{Conformal predictors}: This method predicts based on the similarity of a sample with previous queried samples.
    \item \textbf{Query by committee}: In this strategy, different models are trained, and the samples that most models, called the "committee", disagree have the potential to be queried.
\end{itemize}

As a solution, we propose a heuristic approach, presented in Algorithm~\ref{alg2} for selecting query sample pairs.
The proposed heuristic approach aims to select the most diverse concepts to compare at first and gradually move to similar ones to reduce the search space.
For this purpose, we partition the concepts into clusters based on different similarity measures.
We use semantic and lexical similarity as the features; a similar measure is proposed by~\cite{falke2019automatic} for grouping similar concepts in the process of making a concept map. 
These features include normalized Levenshtein distance, Jaccard coefficient between stemmed content words, semantic similarity based on Latent Semantic Analysis~\cite{deerwester1990indexing}, WordNet~\cite{miller1990introduction}, and word embedding~\cite{mikolov2013distributed}.
Then we model the similarity as a binary classification using logistic regression such that a positive classification, $y= 1$, means that concepts are co-referent. The function is defined as:
\begin{equation}
    P(y= 1|c_1,c_2,\theta) = \mathrm{Sigmoid} (\theta^T\delta(c_1,c_2))
\end{equation}
where $\delta(c_1,c_2)$ are the features, $\theta$ the learned parameters, and the Sigmoid function is defined as:
\begin{equation}
S_{\theta}(z)=(\frac{1}{1+{e}^{\theta(1-z)}})
\end{equation}

Based on the similarity of two concepts, we use an integer linear programming (ILP) function to find an optimized partitioning schema that maximally agrees with the pairwise classifications proposed by~\cite{barzilay2006aggregation} and is transitive due to the constraints.
Let $x_{ij}\in \{0,1\}$ be a binary value represents the co-reference of concepts $(c_i,c_j)$ and $p(c_i,c_j)$ the co-reference probabilities of them.
The goal is to optimize the objective function in Eq.~\ref{max} using a greedy local search to partition similar concepts.
\begin{equation}
\begin{split}
&\sum_{\substack{c_i,c_j \in C^2}} p(c_i,c_j)x_{ij}+ (1-p(c_i,c_j)) (1-x_{ij})\\
& s.t. \hspace{2mm} x_{ik} \geq x_{ij}+x_{jk}-1 \hspace{2mm} \forall i,j,k \in [1,..,|C|]\hspace{2mm} and \hspace{2mm} i \neq j \neq k
\end{split}
\label{max}
\end{equation}

After partitioning similar concepts, in each iteration (the number of iteration is equal to the query budget), we select two concepts in different partitions.
These concepts are selected based on the trained similarity measure and by minimizing the similarity of concepts chosen and previously queried pairs to gain maximum information. 
\begin{algorithm}[t]
    \caption{Modeling User Preference.}\label{alg2}
    \begin{flushleft}
    \hspace*{\algorithmicindent} \textbf{Input}: Concepts, learning rate ($\gamma_1$), query budget $t$\\
    \hspace*{\algorithmicindent} \textbf{Output}: Concept Ranker Function (R). 
    \end{flushleft}
    \begin{algorithmic}[1]
    \Procedure{Modeling User Preference.}{}
    \BState \emph{While $i=0,...,t_1$}:
    \State $\textit{$(c_{1i},c_{2i})$}  \gets \textit{Select a pair based on Eq.~\ref{max}}.$
    \State $\textit{$p(c_{1i},c_{2i})$} \gets \textit{Query user for the feedback.}$
    \State $\textit{$w_{i+1}=w_i+\gamma_1 \frac{\delta J_x(w)}{w}$ based on Eq.~\ref{objectiveTR}.}$
    \EndProcedure
    \BState \emph{end while}
    \State $\textit{Return Ranker Function (R)}$
    \end{algorithmic}
\end{algorithm}

\subsection{The summarizer}
The user preference extractor's output is the ranking function that estimates each concept's importance based on the users' feedback.
The summarizer is responsible for making desired summaries for users based on their given preferences.
Our summarizer consists of three phases: i) A summary generator, ii) An inverse reinforcement learner for evaluating the generated summaries based on the expert's evaluation history, and iii) A reinforcement learner to learn how to generate the desired summary for the user, as discussed below.

\subsubsection{The Summary Generator}
After learning the importance of concepts for a user, $R$ function, we construct summaries that are more likely the desired summary for the user to reduce the summary search space.
Let $C(x)$ be the set of concepts in a given set of source documents $x$, $p_{c_i}$ the presence of the concept $c_i$ in the resulting summary that ${c_i}$ belongs to this sentence, $w_i$ the concept's weight (importance), $l_j$ the length of sentence $j$, $p_{s_j}$ the presence of sentence $j$ in the summary, and $L$ the summary length constraint defined by the user.
Based on these definitions, we formulate the following optimization function using Integer Linear Programming (ILP), which selects sentences with important concepts based on user's feedback, defined as:
\begin{equation} \label{sumGenEq}
\begin{split}
  &  max \sum_{i} w_ip_{c_i}\hspace{2mm} where \hspace{2mm}  \forall i\in [1,..,|C|] \hspace{2mm} and \hspace{2mm} \forall c_i \in s_j \sum_{j} l_jp_{s_j}<L 
\end{split}
\end{equation}

Weights are based on the $R$ function learned in the previous part.
Then a summary pool is made using the above function.
To make a diverse summary pool, among top score summaries, according to Eq.~\ref{sumGenEq}, we select summaries where they are not redundant.
The redundancy is defined as the similarity of sentences inside a summary without considering the user's mentioned concepts divided by the summary length.
Document summarization is then formulated as a sequential decision-making problem, solving by a proposed reinforcement learning (RL) algorithm.  
In the following, we explain the problem definition.

\subsubsection{Problem Definition}
We formulate summarization as a discrete optimization problem inspired by the APRIL approach~\cite{gao2018april}. 
Let $Y_x$ indicate the set of all extractive summaries for the document cluster $x$ and $y_x \in Y_x$ is a potential summary for document cluster $x$.
An input can be either a single document or a cluster of documents on the same topic.
The summarization task is to map each input $x$ to its best summary in $Y_x$ for the learned preference ranking function.
EMDS can be defined as a sequential decision-making problem, sequentially select sentences from the original documents and add them to a draft summary.
Therefore, it can be defined as an episodic MDP (Markov Decision Process) problem described below.

An episodic MDP is a tuple $(S,A,P,R,T)$ where $S$ is the set of states, $A$ is the set of actions, $P:S\times A \times S \xrightarrow{} \mathbbm{R} $ is the transition function, $R(s,a)$ is the reward action performing an action($a$) in a state ($s$) and $T$ is the set of terminal states.
In the EMDS context, as defined in~\cite{gao2019reward}, a state is a draft summary and $A$ includes two types of actions: concatenate a new sentence to the current draft summary or terminate the draft summary construction. 
The reward function $R$ returns an evaluation score of the summary once the action terminates is performed; otherwise, it returns $0$ because the summary is still under construction and hence can not be evaluated.
A policy $\pi(s,a): S \times A \xrightarrow{} R$ in an MDP defines how actions are selected in state $s$.

Episodic MDP for modeling document summarization has two components: i) reward: what is defined as a good summary, ii) policy: how to select sentences (actions) to maximize the rewards.
State-of-the-art summarization approaches are divided into two categories:  cross-input paradigm and  input-specific~\cite{gao2019reward}.
The former employs reinforcement learning algorithms such that the agent interacts with a ground-truth reward oracle over multiple episodes to learn a policy that maximizes the accumulated reward in the episode. 
The learned policy is used to apply on unseen data at test time for generating summaries.
However, learning such a cross-input policy requires considerable time, data, and parameter tuning due to the vast search spaces and delayed rewards.
On the other side, learning input-specific RL policies is a more efficient alternative that agent interacts to learn a policy specifically for the given input without requiring parallel data or reward oracle.
However, they depend on handcrafted rewards, challenging to design to fit all inputs~\cite{gao2019reward}.

\textit{SumRecom} takes advantage of two categories of cross-input and input-specific reinforcement learning.
First, it learns a cross-input reward oracle at training time and then uses the learned reward to train an input-specific policy for each input at test time, as discussed below.

\subsubsection{The Reward Learner}
\textit{SumRecom} is inspired by inverse reinforcement learning~\cite{abbeel2004apprenticeship}, where instead of the policy, it first learns the reward utilizing an \textit{expert demonstrator} to present optimal trajectories.
The demonstrator is a domain expert who can evaluate two summaries based on the users' given feedback.
The demonstrator can be another RL agent trained to become an advisor for other learner-agents or a human.
To approximate the ground-truth reward oracle from “weak supervisions”, numeric scores that indicate the quality of the summary and preferences over summary pairs are used as humans reliably provide such judgments~\cite{kreutzer2018reliability}.
In practice, leveraging preference learning reduces the cognitive load and, consequently, reduces the inevitable noise in evaluating summaries.
In this paper, to evaluate \textit{SumRecom}, it learns from preference-based (pairwise) oracles that provide preferences over summary pairs and point-based oracles that provide point-based scores for summaries.
In both cases, a summary is selected from the summary-pool created by the summary generator component.
Then, the summaries are queried to the demonstrator for evaluation.
In point-based, we draw $L$ sample outputs from the summary pool without replacement.

We use our previous work strategy~\cite{ghodratnama2020extractive}, ExDos, to evaluate the summaries based on three measures: coverage, salience, and redundancy.
ExDoS combines both supervised and unsupervised algorithms in a single framework and an interpretable manner for document summarization purpose.
ExDoS iteratively minimizes the error rate of the classifier in each cluster with the help of dynamic local feature weighting. Moreover, this approach specifies the contribution of features to discriminate each class.
Therefore, in addition to summarizing text, ExDoS is also able to measure the importance of each feature in the summarization process. 
In each iteration, summaries with the most difference with previously selected samples are chosen for being queried.
The same approach is also selected for preference-based summaries.
Then, we query their score values ($V$) from the oracle and use a regression algorithm to minimize the averaged mean squared error (MSE) between $V$ and the approximate value $V^*$ where the loss function is:
\newcommand{\Lagr}{\mathcal{L}}
\begin{equation}
\label{los}
    \Lagr^{MSE}= \frac{1}{L}\sum_{i=1}^{L}(V^*-V)^2
\end{equation}
In pairwise , we denote the collected preferences by $P_s=\{p(y_{11},y_{12}),...,p(y_{1l},y_{2l})\}$ where $y$ denotes the summary and $l$ sample pairs are queried.
Then the procedure is the same as in Eq.~\ref{objectiveTR} using cross-entropy loss function defined as:
\begin{equation} \label{losspresum}
\begin{split}
  \Lagr^{CE}= & -\sum_{i \in l}[p(y_{1i},y_{2i})\log H(y_{1i},y_{2i};w)+\\
 &  [p(y_{2i},y_{1i})\log H(y_{2i},y_{1i};w)))]
\end{split}
\end{equation}
where
\begin{equation} \label{eq21}
 H(y_{1i},y_{2i};w)= \frac{1}{1+\exp [V^{*}{(y_j;w)}-V^{*}{(y_i;w)}]}
\end{equation}

The output is the ranked function, $V$, which demonstrates each summary's reward compared to the others.
Because such a demonstrator is hardly available in practice, \textit{SumRecom} leverages an approximate function to learn the reward to the summaries generated in the previous step explained in the experiment section.
\begin{algorithm}[t]
    \caption{Summarizer}\label{alg3}
    \begin{flushleft}
    \hspace*{\algorithmicindent} \textbf{Input}: Concept ranker (R), learning rate ($\gamma_2$). \\
    \hspace*{\algorithmicindent} \textbf{Output}: Optimal Summary for a user (S). 
    \end{flushleft}
    \begin{algorithmic}[1]
    \Procedure{Summarizer.}{}
    \BState \emph{Summary Generator}:
    \State $\textit{Summaries} \gets \textit{Generating Summaries using Eq.~\ref{sumGenEq}.}$
    \BState \emph{Reward Learner}:
    \BState \emph{While $i=0,...,t_2$}:
    \State $\textit{$(y_{1i},y_{2i})$}  \gets \textit{Select a pair based on Eq.~\ref{losspresum} from summaries.}$
    \State $\textit{$p(y_{1i},y_{2i})$} \gets \textit{Query user for the feedback.}$
    \State $\textit{$w_{t+1}=w_t-\gamma_2 \frac{ \Lagr^{CE}(w)}{\delta w}$ based on Eq.~\ref{los} or Eq.\ref{losspresum}.}$
    \BState \emph{end while}
    \BState \emph{Policy Learner}:
    \State $\pi^* = arg max R^{RL}(\pi|x) = arg max \sum_{y in Y}\pi(y)V(y)$
    \EndProcedure
    \end{algorithmic}
\end{algorithm}

\subsubsection{The Policy Learner}
The goal of policy learning is to search for optimal solutions in Markov Decision Processes (MDPs).
We model the summarization problem as an episodic MDP, meaning that each action's reward is equal to zero if the state is not terminate.
At each step, the agent can perform either of the two actions: add another sentence to the summary or terminate it.
The immediate reward function $R(s,a)$ assigns the reward if $s$ is the terminate state.
The reward in SumRecom is the learned expert's reward, $V$.
In EMDS, a policy $\pi$ defines the strategy to add sentences to the draft summary to build the summary for the user.
\textit{SumRecom} defines $\pi$ as the probability of choosing a summary of $y$ among all summaries $Y$ denoted as $\pi(y)$. 
Therefore, the optimal policy, $\pi^*$, is the function that finds the desired summary for a given input based on the user's feedback.
The expected reward of performing proper policy $\pi$ is defined as:
\begin{equation}
    R^{RL}(\pi|x)= \mathbbm{E}_{y \in Y}R(y)= \sum_{y\in Y} \pi(y)R(y)
\end{equation}
where $R(y)$ is the reward for selecting summary $y$ in document cluster x.
In our problem, the reward is the ranker approximated by the domain expert, $V$.
Therefore, the accumulated reward to be maximized in our problem is equal to:
\begin{equation}
    R^{RL}(\pi|x)= \sum_{y\in Y} \pi(y)V(y)
\end{equation}

The goal of MDP is to find the optimal policy $\pi^*$ that has the highest expected reward:
\begin{equation}
\pi^* = arg max R^{RL}(\pi|x) = arg max \sum_{y \in Y}\pi(y) V(y)
\end{equation}
We use the \textit{linear Temporal Difference (TD)} algorithm to obtain  $\pi^*$.
The summarizer algorithm is explained in Algorithm~\ref{alg3}.

\section{Evaluation}
\label{Evaluation}
In this section, we present the experimental setup for implementing and assessing our summarization model's performance.
We discuss the datasets, give implementation details, and explain how system output was evaluated.

\subsection{Datasets}
We evaluated \textit{SumRecom} using three commonly employed benchmark datasets from the Document Understanding Conferences (DUC) \footnote{Produced by the National Institute Standards and Technology (https://duc.nist.gov/)}.
The DUC dataset details are described in Table~\ref{tab:dataset}.
Each dataset contains a set of document clusters accompanied by several human-generated summaries used for training and evaluation. 

\subsection{Evaluation Metric}
We evaluate the quality of summaries using $ROUGE_N$ measure~\cite{lin2004rouge}\footnote{We run ROUGE 1.5.5: http://www.berouge.com/Pages/defailt.aspx with parameters -n 2 -m -u -c 95 -r 1000 -f A -p 0.5 -t 0} defined as:
\begin{table}[t]
\centering
  \centering \caption{Dataset description, indicating number of documents, number of document clusters, and average number of sentences in each document.}
  \vspace{1mm}
  \label{tab:dataset}
  \tabcolsep=0.11cm
  \begin{tabular}{|c|c|c|l|}
    \hline
    Dataset & Doc-Num & Cluster-Num & Sentence\\
   \hline
    DUC1 &30&308&378\\
    \hline
    DUC2 &59&567&271\\
    \hline
    DUC4 &50&500&265\\
    \hline
\end{tabular}
\end{table}

\begin{equation}
ROUGE_N= \frac{\sum_{S\in \{Reference Summaries\}} \sum_{gram_n \in S}{Count_{match}(gram_n)}}{\sum_{S\in \{Reference Summaries\}} \sum_{gram_n \in S}{Count(gram_n)}}
\end{equation}

The three variants of ROUGE (ROUGE-1, ROUGE-2, and ROUGE-L) are used.
We used the limited length ROUGE recall-only evaluation (75 words) to compare DUC dataset to avoid being biased. 


\subsubsection{Results and Analysis}
First, we explain the evaluation settings, and then we discuss the results and analysis.
\textit{SumRecom} is evaluated from different evaluation aspects, including:
\begin{itemize}
    \item The impact of different features in approximating preference learning algorithms used in this paper, including concept and summary preferences.
    \item The use of different strategies for active learning.
    \item The effect of different concept units: unigram, bigram, and sentence.
    \item The role of the query budget in both concepts and summary preferences.
    \item The quality of produced summaries.
    \item A human study to evaluate \textit{SumRecom} through users' lens.
    \item The effect of different values for parameters. 
    \item An ablation study to evaluate different components of the proposed framework.
\end{itemize}
We have simulated users in all experiments except the one proposed for human evaluation.

\begin{figure}
\centering
    \includegraphics[width=0.7\linewidth]{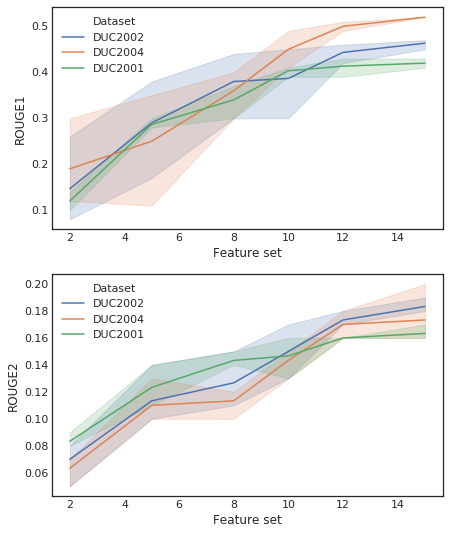}
    \caption{Features for estimating the ranker function for preferred concepts.}
    \label{featureAnalysis}
\end{figure}
\textbf{Feature Analysis.~\footnote{We use nltk library, which has the ngrams function that returns a generator of n-grams given a tokenized sentence for experiments that mentions 1-gram, 2-gram, and sentence. In another attempt on human evaluation, we used Open Information Extraction~\cite{yates2007textrunner}, an approach that extracts binary propositions from the text.}} Before evaluating the effect of concept preference in summarization, we require to explain the ground-truth concept ranker function ($U$) and the approximate function ($U^*$).
The ground-truth concept ranker function ($U$) indicates the importance of each concept. 
We define a predefined list of preference over concepts and the ground-truth concept ranker value for ten clusters to simulate the users' preferences.
For estimating the approximate function ($U^*$), we define a linear model $U^*(c)=W^T\phi(c)$ where $\phi$ are the features.
To this end, a set of features that their importance is validated in our previous work~\cite{ghodratnama2020extractive}, including surface-level and linguistic-level features are used.
Surface-level features include frequency-based features (TF-IDF, RIDF, gain, and word co-occurrence), word-based features (uppercase word and signature words), similarity-based features (Word2Vec and Jaccard measure), sentence-level features (position, length cut off and length), and Named Entity.
Linguistic features are made based on the semantic graph.
These features include the average weights of connected edges, the merge status of a sentence as a binary feature, the number of concepts merged with a concept, and the number of concepts connected to the concept.

\begin{figure}[t]
\centering
    \includegraphics[width=0.75\linewidth]{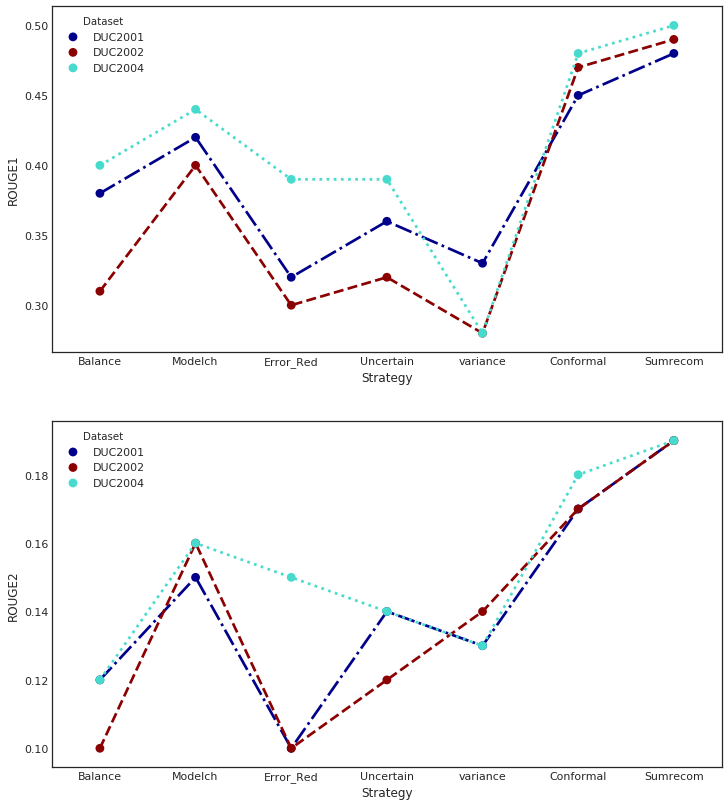}
    \caption{Comparing different strategies used in active learning.}
    \label{fig:strategies}
\end{figure}

We defined different combinations of features, $\{2,5,8,10,12,15\}$, starting from the most critical one based on the importance estimated in our previous work~\cite{ghodratnama2020extractive}. 
We repeated the experiments for $10$ cluster documents.
The results of ROUGE1 and ROUGE 2 are reported in Figure~\ref{featureAnalysis}.

As results show, the performance increased by adding more features; however, the last set of features did not significantly impact ROUGE values.
However, it is worth mentioning that adding more features increases complexity.
To simulate the domain-expert knowledge for evaluating summaries based on the given feedback (reward), we model the ground-truth summary reward ($V$) based on the three measures, including ROUGE1, ROUGE2, and the redundancy, defined as:
\begin{equation}
\label{reward}
    V=\alpha ROUGE1+ \beta ROUGE2 - \gamma Redundancy
\end{equation}

ROUGE1 and ROUGE2 are arguably the most widely used metrics to approximate human evaluation of summary quality.
Redundancy is defined as the similarity of sentences inside a summary without considering the user's mentioned concepts, divided by the summary length.
\begin{figure}[t]
\centering
    \includegraphics[width=0.75\linewidth]{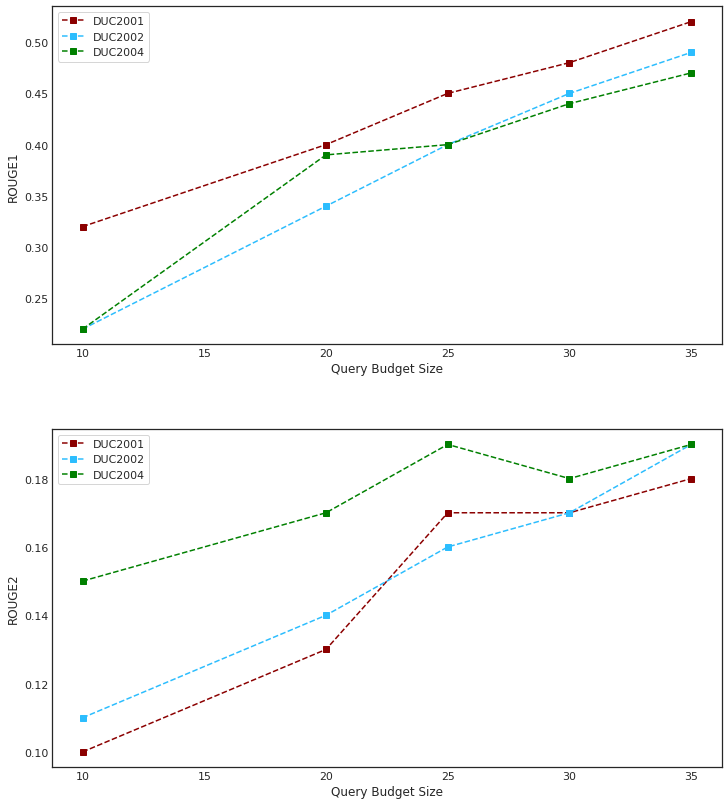}
    \caption{The effect of query budget size in quality of summaries.}
    \label{querybudget}
\end{figure}
To approximate the ground-truth reward function, we employ a linear function as $V^*=w^T\lambda(y)$, where $\lambda(y)$ is the combination of features used for the concepts plus ROUGE1 and ROUGE2.
The results follow the same trend in Figure~\ref{featureAnalysis} except that adding ROUGE1 and ROUGE2 as the feature set improves the performance of final summaries by the average of $1.13$ times.\\

\begin{figure}
\centering
    \includegraphics[width=0.6\linewidth]{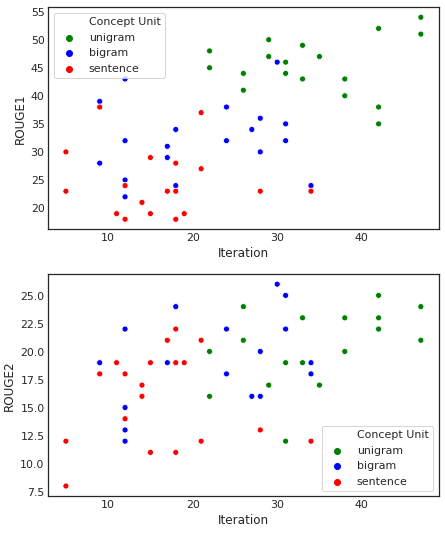}
    \caption{Analysis of effect of concept unit including unigram, bigram and sentence.}
    \label{conceptunit}
\end{figure}

\begin{table}
\centering
  \caption{Comparing SumRecom on DUC1 dataset.}
  \vspace{1mm}
  \label{tab:data1}
  \begin{tabular}{|l|c|c|c|}
    \hline
    Model &ROUGE-1 &ROUGE-2 &ROUGE-L\\
    \hline
    APRIL &0.325  &0.070 &0.26\\
    \hline
    SPPI  &0.232  &0.068 &0.259\\
    \hline 
    SumRecom &0.341 &0.078 &0.28\\
 \hline
\end{tabular}
\end{table}
\begin{table}
\centering
  \caption{Comparing SumRecom on DUC2 dataset.}
    \vspace{1mm}
  \label{tab:data2}
  \begin{tabular}{|l|c|c|c|}
     \hline
    Model &ROUGE-1 &ROUGE-2 &ROUGE-L\\
    \hline
     APRIL &0.351 &0.078 &0.279\\
      \hline
     SPPI &0.350 &0.077 &0.278\\
      \hline
     SumRecom &0.372 &0.083 &0.333\\
   \hline
\end{tabular}
\end{table}
\begin{table}
\centering
  \caption{Comparing SumRecom on DUC2004 dataset.}
    \vspace{1mm}
  \label{tab:data3}
  \begin{tabular}{|l|c|c|c|}
     \hline
    Model & ROUGE-1 & ROUGE-2 & ROUGE-L\\
    \hline
     APRI L&0.373 &0.093 &0.293\\
      \hline
     SPPI &0.372 &0.093 &0.293\\
      \hline
     SumRecom &0.382 &0.094 &0.301\\
   \hline
\end{tabular}
\end{table}
\begin{figure}
\centering
    \centerline{\includegraphics[width=1.1\textwidth]{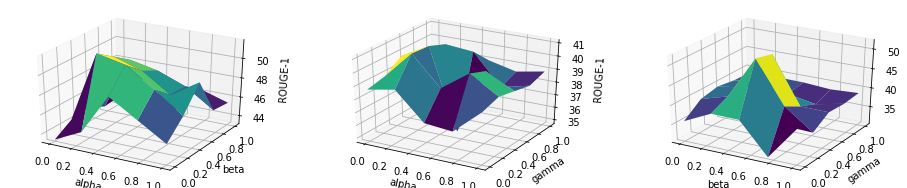}}
    \caption{Evaluation of different parameters and the ROUGE value for $\alpha$, $\beta$, and $\gamma$, which are hyper-parameters of the ground-truth reward function.}
    \label{parameter}
\end{figure}
\begin{table}
\centering
  \caption{Overview of the parameters used in simulation experiments. }
  \label{tab:parameter}
  \begin{tabular}{|c|c|l|}
     \hline
    Parameter & Description & Value\\
    \hline
    L & user input &summary length\\
     \hline
    $\alpha$ &0.8 &ROUGE1 coefficient for ground truth reward in Eq.~\ref{reward}\\
     \hline
    $\beta$ &0.5  &ROUGE2 coefficient for ground truth reward in Eq.~\ref{reward}\\
     \hline
    $\gamma$ &0.25 &redundancy coefficient for ground truth  reward in Eq.~\ref{reward}\\
     \hline
    $\gamma_1$ &0.001 &learning rate for concept preference in Eq.~\ref{objectiveTR}\\
     \hline
    $\gamma_2$ &0.005 &learning rate for summary preference in Eq.~\ref{los}\\
     \hline
\end{tabular}
\end{table}

\begin{table}
\centering
  \caption{Comparing SumRecom, SumRecom-AC, and SumRecom-PR on DUC1 dataset.}
   \label{AblPreference1}
  \begin{tabular}{|l|c|c|c|}
     \hline
    Model & ROUGE1 & ROUGE2 & ROUGEL\\
    \hline
     SumRecom-AC &0.103 &0.031 &0.140\\
      \hline
     SumRecom-PR &0.112 &0.001 &0.129\\
      \hline
     SumRecom &0.341 &0.078 &0.28\\
   \hline
\end{tabular}
\end{table}

\begin{table}
\centering
  \caption{Comparing SumRecom, SumRecom-AC, and SumRecom-PR on DUC2 dataset.}
   \label{AblPreference2}
  \begin{tabular}{|l|c|c|c|}
     \hline
    Model & ROUGE1 & ROUGE2 & ROUGEL\\
    \hline
     SumRecom-AC &0.190 &0.018 &0.132\\
      \hline
     SumRecom-PR &0.157 &0.021 &0.198\\
      \hline
     SumRecom &0.572 &0.083 &0.333\\
   \hline
\end{tabular}
\end{table}

\begin{table}
\centering
  \caption{Comparing SumRecom, SumRecom-AC, and SumRecom-PR on DUC4 dataset.}
   \label{AblPreference4}
  \begin{tabular}{|l|c|c|c|}
     \hline
    Model & ROUGE1 & ROUGE2 & ROUGEL\\
    \hline
     SumRecom-AC &0.111 &0.033 &0.143\\
      \hline
     SumRecom-PR &0.200 &0.021 &0.182\\
      \hline
     SumRecom &0.382 &0.094 &0.301\\
   \hline
\end{tabular}
\end{table}

\begin{table}
\centering
  \caption{Comparing SumRecom and SumRecom-GE.}
   \label{tge}
  \begin{tabular}{|l|l|l|l|l|l|l|}
    \hline
    \multirow{2}{*}{Dataset} &
      \multicolumn{2}{c}{SumRecom} &
      \multicolumn{2}{c|}{SumRecom-GE} \\
    & R1 & R2 & R1 & R2 \\
    \hline
    DUC1 & 0.341 & 0.078 & 0.222 & 0.021 \\
    \hline
    DUC2 & 0.572 & 0.083 & 0.189 & 0.047  \\
    \hline
    DUC4 & 0.382 & 0.094 & 0.109 & 0.510  \\
    \hline
  \end{tabular}
\end{table}

\textbf{Active Learning Strategy Analysis.} To evaluate the impact of the proposed heuristic for active learning, we compared \textit{SumRecom} with six different strategies explained in section~\ref{sec:activelearn}.
Ten clusters from each dataset is used for experiments.
ROUGE1 and ROUGE2 results for each strategy are reported in Figure~\ref{fig:strategies}, proving the supremacy of the proposed heuristic approach for our problem. 
As the pictures reveal, the \textit{conformal approach} acts approximately similar to the proposed heuristic function.
Besides, the \textit{change model} acts better than others in both cases approving that selecting the most different concepts results in better summaries in this problem.\\

\textbf{Query Budget Analysis.} We also measure the effectiveness of the users' query budget size in the process.
We selected the query size among the selection of $\{10,15,20,25,30,35\}$, demonstrating the user's number of feedback.
The results are reported in Figure~\ref{querybudget}.
As expected, by increasing the number of feedback, the ROUGE score increases significantly.
However, the difference rate decreases through the process.\\

\textbf{Concept Unit Analysis.} In another experiment, to evaluate the concept unit's impact, the preference unit showed to users for feedback; we analyzed three-unit measures: uni-gram, bi-gram, and sentence.
Then we evaluate the number of iterations required to extract feedback for reaching the upper-bound using each concept unit.
As Figure~\ref{conceptunit} shows, unigram-based feedback requires significantly more feedback to converge, while sentence need less feedback to converge.\\

\textbf{Summary Evaluation.} To evaluate the coverage aspect of summaries generated by \textit{SumRecom}, we used the reference summaries such that the mentioned concepts which exist in reference summaries get the maximum score by the ranked function.
Besides, the reward function is the ROUGE score by comparing to reference summaries.
We compared SumRecom to state-of-the-art strength competitors including \textit{APRIL} and \textit{SPPI} on three bench mark dataset reported in Table~\ref{tab:data1}, Table~\ref{tab:data2}, and Table~\ref{tab:data3}.
The results show the supremacy of \textit{SumRecom}.
Besides, the main goal of \textit{SumRecom} is to help users make their desired summary with the least cognitive load.
Therefore, in the next step, we conducted a study to evaluate the cognitive load of users.\\

\textbf{Human Analysis.} The goal of \textit{SumRecom} is to help users make the desired summary with low cognitive load.
Therefore, we conducted two human experiments to evaluate the model.
We hired fifteen Amazon Mechanical Turk (AMT) workers to perform the tasks without any specific prior background required.
Ten document clusters were randomly selected from the DUC datasets. 
Each participant was presented with five random documents to avoid any subjects' bias and was given two minutes to read each article.
To make sure human subjects understood the study's objective, we asked workers to complete a qualification task first.
They were required to write a summary of their understanding.
We also manually removed spam answers from our results.
Spams are defined based on the qualification task and the response time (very short answering time is unacceptable as it proves random or imprecise answers).

In the first experiment, participants were asked to define their preferences by comparing some concepts.
The generated summary based on the given feedback and four other general summaries were shown to them, and they were asked to choose their preferred summary.
$83\%$ of participants selected the generated summary produced by \textit{SumRecom}.
Then, they were asked to define their satisfaction level to evaluate the produced summary based on their given feedback by assigning a rate between zero and ten.
The average rate of summaries produced by \textit{SumRecom} were $8.2$, demonstrating a reasonable level of satisfaction.
To evaluate the effect of cognitive load in the proposed approach compared to state-of-the-art techniques that compare the summaries (APRIL), we ask participants to select preference over summaries and their average response time in another experiment.
On average, the proposed approach had $2.3$ times less response time (13 seconds).

In the second experiment, to assess whether users can find their desired information, they were asked to answer a given question about each topic by selecting an answer among the given potential answers.
These questions are defined by the authors and cover both specific and general information.
Their level of confidence in answering questions and their answers were recorded.
An evaluator assessed their accuracy in answering questions.
Among the fifteen workers, 86.67\% were completely confident in their answers.
However, 80\% percent answered accurately.\\

\textbf{Parameter Analysis.} As in other parametric models, \textit{SumRecom} has some hyper-parameters needed to be tuned.
A wide range of these values is tested, and the correlation between them is analyzed.
As an example, Figure~\ref{parameter} evaluates different parameters and the ROUGE value for $\alpha$, $\beta$, and $\gamma$, which are hyper-parameters of the ground-truth reward function.
To ease the reading and help researchers replicate the proposed algorithm, we summarize the parameters with their description we used in our simulation experiments in Table~\ref{tab:parameter}.\\

\textbf{Ablation Study.} 
To evaluate each component's incremental contribution in our proposed framework, we run ablation studies comparing our model ablations against each other.

First, we evaluate the \textit{preference extractor} part.
Therefore, we remove active learning by selecting random pairs from the concept database, called \textit{SumRecom-AC}.
Then the whole preference learner part was removed, called \textit{SumRecom-PR}.
The ROUGE scores of these approaches comparing with reference summaries after ten complete runs are averaged and compared in Table~\ref{AblPreference1}, Table~\ref{AblPreference2}, and Table~\ref{AblPreference4}, respectively.
The results clearly show the effect of both \textit{active learning} and \textit{preference learning}.

In another experiment to evaluate the role of the learning process, the summaries generated by \textit{summary generator} are considered as output, called \textit{SumRecom-GE}.
The results are reported in Table~\ref{tge} and are the average of ten produced summaries.

\section{Conclusion and Future Work}
\label{conclusion}
We propose a summary recommendation framework that interactively learns to generate personalized summaries based on users' feedback.
We took a step further from the current state-of-the-art summarization approaches by considering three research questions in this work:
i) Can user preferences over concepts provide personalized summaries that reflect users' interests with less cognitive load?
ii) Can domain-expert knowledge be embedded in the learning process?
iii) How can user preferences and domain-experts experience be modeled as a reinforcement learning algorithm to generate desired summaries for users automatically?

Employing users' feedback and domain expert's knowledge in a reinforcement learning algorithm demonstrates that SumRecom could generate desired summaries for users.
However, there still many challenges needs to overcome.
Therefore, many future directions are possible for future work.
First, capturing users’ interests is a significant challenge in providing effective personalized information. 
The reason is that users are reluctant to specify their preferences as entering lists of interests may be a tedious and time-consuming process. 
Besides, people’s interests are not static and change over time that should be taken into account.
Therefore, techniques that extract implicit information about users’ preferences are the next step for making effective personalized summaries.
Another potential direction is to use human feedback history to provide personalized summaries on new domains using transfer learning.\\

\bibliographystyle{unsrt}  
\bibliography{references.bib}

\end{document}